\documentclass[12pt]{article}
\pdfoutput=1

\usepackage[utf8]{inputenc}
\usepackage[margin=1in]{geometry}
\usepackage[tbtags]{amsmath} 				
\usepackage{amssymb, mathrsfs}			
\usepackage{physics}
\usepackage{graphicx}			
\usepackage{tensor}				
\usepackage[makeroom]{cancel}	
\usepackage{bm}		
\usepackage{bbold}
\usepackage[
      colorlinks=true,
      linkcolor=blue,
      urlcolor=blue,
      filecolor=black,
      citecolor=red,
      pdfstartview=FitV,
      pdftitle={},
      pdfauthor={},
      bookmarksopen=true
      ]{hyperref}

\def\cA{{\cal A}}
\def\cB{{\cal B}}

\def\cG{{\cal G}}

\def\cN{{\cal N}}

\def\Im{{\rm Im \,}}

\usepackage{sectsty}
\sectionfont{\large}

\renewcommand{\thefootnote}{\fnsymbol{footnote}}
\renewcommand{\thanks}[1]{\footnote{#1}}
\newcommand{\starttext}{
\setcounter{footnote}{0}
\renewcommand{\thefootnote}{\arabic{footnote}}}
\renewcommand\({\begin{equation}}		
\renewcommand\){\end{equation}}
\renewcommand{\epsilon}{\varepsilon}	

\numberwithin{equation}{section} 		

\numberwithin{equation}{section}

\renewcommand{\Im}{\operatorname{Im}}

\long\def\symbolfootnote[#1]#2{\begingroup%
\def\thefootnote{\fnsymbol{footnote}}\footnote[#1]{#2}\endgroup}

\begin{document}
\setlength{\baselineskip}{16pt}

\starttext
\setcounter{footnote}{0}

\begin{flushright}
\today
\end{flushright}

\bigskip

\begin{center}

{\Large \bf   A Penrose limit  for type IIB $AdS_6$ solutions}

\vskip 0.4in

\vskip 0.4in

{\large  Michael Gutperle  and Nicholas Klein  }

\vskip 0.2in

{\sl Mani L. Bhaumik Institute for Theoretical Physics}\\
{\sl Department of Physics and Astronomy }\\
{\sl University of California, Los Angeles, CA 90095, USA} 

\bigskip

\end{center}
 
\begin{abstract}
\setlength{\baselineskip}{16pt}

In this paper  a Penrose limit  is constructed for type IIB $AdS_6\times S^2\times \Sigma$ supergravity solutions. These solutions are  dual to five dimensional SCFTs \ related to (p,q) five brane webs,  which can often be described in terms of long quiver 
gauge theories. The null geodesic from which the Penrose limit is constructed is localized at a unique point  on the two dimensional Riemann surface  $\Sigma$, where the $AdS_6$ and $S^2$ metric factors are extremal.  The resulting pp-wave spacetime takes a  universal form. The  world sheet action of the Green-Schwarz string  is quadratic in the light  cone gauge  and the spectrum  of  string excitations is obtained.
\end{abstract}

\setcounter{equation}{0}
\setcounter{footnote}{0}

\newpage

\section{Introduction}

The quantization of  superstrings in the presence of   Ramond-Ramond fields is  a technically challenging problem and  not solved  in general. This constitutes a challenge to go beyond the  semi-classical supergravity approximation in  many examples of AdS/CFT.  Instead of solving the general problem one approach is to consider limits or deformations of the supergravity solution which may lead to a simpler quantization  problem. One such limit is given  by taking the Penrose limit of an $AdS_p\times S^q$  background. The Penrose limit \cite{Penrose:1976A} (see also \cite{Gueven:1987ad,Sfetsos:1993rh,Sfetsos:1994fc}) corresponds to zooming in on the close vicinity of a null geodesic in the spacetime and produces a plane wave geometry.
In the case of the Penrose limit of the type IIB $AdS_5\times S^5$ solution, where the null geodesic sits at the center of $AdS_5$ and on a great circle of $S^5$, one obtains  a maximally supersymmetric plane wave  \cite{Figueroa-OFarrill:2001hal,Blau:2002dy,Blau:2001ne}. It was proposed in  \cite{Berenstein:2002jq} that on the field theory side this limit corresponds to singling out a special subset of CFT operators where both the conformal dimension $\Delta$ and a $U(1)$ R-charge $J$  are taken to be of order $\sqrt{N}$, with the difference  $\Delta-J$ finite, as $N\to \infty$.  One important feature of this type IIB  plane wave background is that the Green-Schwarz string can be quantized exactly \cite{Metsaev:2001bj,Metsaev:2002re,Mizoguchi:2002qy}. The world sheet theory in the light cone gauge corresponds to the action of eight  massive bosons and fermions. Furthermore the machinery of light-cone string field theory can be used to calculate string interactions and compare the results to field theory \cite{Constable:2002hw,Beisert:2002bb,Spradlin:2002ar,Constable:2002vq}.  For an incomplete list discussing Penrose limits of other supergravity backgrounds see \cite{Nishioka:2008gz,Gomis:2002km,Cvetic:2002si,Cvetic:2002hi,Gursoy:2002tx}.

The aim of this paper is to study the Penrose limit for a different AdS solution, namely the type IIB  
solutions found in \cite{DHoker:2016ujz,DHoker:2017mds}, which are realized as warped products of $AdS_6\times S^2$ over a Riemann surface $\Sigma$ with boundary. These solutions are supported by R-R and NS-NS three form fluxes. Indeed the data characterizing the solution can be identified with the $(p,q)$ five brane charges of semi-infinite five branes forming a five brane web.

Unlike the type IIB  $AdS_5\times S^5$ solution or the $AdS_{4|7}\times S^{7|4}$ solution of M-theory these backgrounds are warped product geometries and   preserve only sixteen of the thirty-two  supersymmetries of ten dimensional type IIB supergravity. This fact makes finding a suitable Penrose limit more challenging, since the   radii of the $AdS_6$ and $S^2$ vary with the location on the Riemann surface $\Sigma$ and therefore a general null geodesic will also have a nontrivial dependence on the Riemann surface. There is a special point on the Riemann surface, namely the critical point of a function $\cG$. At this point the metric factors of the $AdS_6$ and $S^2$ are extremized with respect to the coordinates on $\Sigma$. We choose the null-geodesic on which the Penrose limit is based to be  localized at this critical point on $\Sigma$.

The structure of the paper is as follows: In section \ref{sec2} we review the type IIB supergravity solutions first found in \cite{DHoker:2016ujz,DHoker:2017mds}  which we use in the rest of the paper. In section \ref{sec3} we  define the null geodesic on which the Penrose limit is based and present the resulting type IIB plane wave background, including all the other bosonic supergravity fields. In section \ref{sec4} we discuss the quantization of the Green-Schwarz superstring action in the light cone gauge  for this background. We discuss our results and possible directions for further research in section \ref{sec5}. Some detailed calculations and supplementary materials are relegated to appendices.

\section{Type IIB $AdS_6$ solutions}
\label{sec2}

Solutions of superstring theories with $AdS_6$  factors are candidates  for holographic  duals of  five  dimensional superconformal field theories  (SCFTs).  The AdS/CFT correspondence is an important tool in studying these theories. Since the  unique superconformal algebra $F(4)$ with an $SO(2,5)$ factor has sixteen fermionic generators \cite{Kac:1977em,Nahm:1977tg,Shnider:1988wh},  the supergravity background preserves  only 16 of the 32 supersymmetries of  maximally symmetric type II or M-theory  vacua. Furthermore the solutions are all realized as  warped products of the $AdS_6$ over a base space. The first such solution was obtained in massive type IIA  \cite{Brandhuber:1999np}. In this paper we will focus on type IIB solutions first constructed in \cite{DHoker:2016ujz,DHoker:2017mds}, which is realized as  a warped product  of $AdS_6 \times S^2$ over a two dimensional Riemann surface $\Sigma$ with boundary.

\begin{align}\label{eqn:ansatzmet}
	ds^2 &= f_6^2 \, ds^2 _{\mathrm{AdS}_6} + f_2^2 \, ds^2 _{\mathrm{S}^2} 
	+ 4\hat \rho^2\, |dw|^2~,
	\end{align}
	where $w$ is a complex coordinate on $\Sigma$ and $ds^2_{AdS_6}$ and $ds^2_{S^2}$ are the line elements for unit-radius $AdS_6$ and $S^2$, respectively\footnote{We denote the metric factor of $\Sigma$ by $\hat \rho$ to avoid confusion with $\rho$ which radial direction of $Ads_6$ used later.}. The $AdS_6$ solutions are defined in terms of locally holomorphic functions $\cA_\pm$ on a Riemann surface $\Sigma$.
The metric functions read
\begin{align}\label{eq:metric-functions}
	f_6^2&=\sqrt{6\cG T}~, & f_2^2&=\frac{1}{9}\sqrt{6\cG}\,T ^{-\tfrac{3}{2}}~, & \hat\rho^2&=\frac{\kappa^2}{\sqrt{6\cG}} T^{\tfrac{1}{2}}~.
\end{align}
and they are expressed in terms of the  holomorphic functions  $\cA_\pm$ as follows
\begin{align}\label{eq:kappa2-G}
	\kappa^2&=-|\partial_w \cA_+|^2+|\partial_w \cA_-|^2~,
	&
	\partial_w\cB&=\cA_+\partial_w \cA_- - \cA_-\partial_w\cA_+~,
	\nonumber\\
	\cG&=|\cA_+|^2-|\cA_-|^2+\cB+\bar{\cB}~,
	&
	T^2&=\left(\frac{1+R}{1-R}\right)^2=1+\frac{2|\partial_w\cG|^2}{3\kappa^2 \, \cG }~.
\end{align}
In order to obtain regular  and geodesically complete solutions one has  to impose that $\cG$ vanishes along the boundary of the Riemann  surface, which implies that the $S^2$ shrinks to zero size  and the  spacetime  closes off. In \cite{DHoker:2017mds} a large class of regular solutions were constructed  by  choosing $\Sigma$ to be the upper half  plane and the holomorphic functions  such that $\partial_w \cA_\pm$ have $L$ simple poles localized on the boundary which is the real line.
\begin{align}
\cA_\pm  &= \cA^0_\pm + \sum_{l=1}^L Z_\pm^l \ln(w-p_l), \quad \sum_{l=1}^L Z_{\pm}^l=0, \quad Z_{\pm}^l = -{\bar  Z}_{\mp}^l 
\end{align}
The  regularity condition translates into  $L$ conditions \cite{DHoker:2017mds}
\begin{align}
\cA^0_+ Z_-^k - \cA^0_- Z^k_+ + \sum_{l=1, l\neq k}^L (Z_+^l Z_-^k-Z_+^k Z_-^l) \ln |p_l-p_k|=0, \quad \quad k=1,2,\cdots, L
\end{align}
The other non vanishing type IIB supergravity fields of the solution are the complex scalar   $B$ which is related to the axion dilaton $\tau= \chi+ i e^{-\phi}$ via $B=(1+i\tau)/(1-i\tau)$ and the complex two form antisymmetric tensor potential $C_{(2)}$
\begin{align}\label{bc2sol}
B &=\frac{\partial_w \cA_+ \,  \partial_{\bar w} \cG - R \, \partial_{\bar w} \bar \cA_-   \partial_w \cG}{
		R \, \partial_{\bar w}  \bar \cA_+ \partial_w \cG - \partial_w \cA_- \partial_{\bar w}  \cG}~,\quad 
	C_{(2)}&=\frac{2i}{3}\left(
\frac{\partial_{\bar w}\cG\partial_w\cA_++\partial_w \cG \partial_{\bar w}\bar\cA_-}{3\kappa^{2}T^2} - \bar{\mathcal{A}}_{-} - \mathcal{A}_{+}  \right){\rm vol}_{S^2}~.
\end{align}
As discussed in \cite{DHoker:2017mds} the residues of $\cA_\pm$ can be identified with the charges of the $(p,q)$ five brane web which realizes the dual SCFT via
\begin{align}
Z_+^ l = {3\over 4} \alpha' (p_l+i q_l ), \quad  l=1, 2,\cdots, L
\end{align} 

\subsection{Explicit examples}
For definiteness we consider two  explicit examples of the regular supergravity solution for which the dual SCFT are well studied, namely the $T_N$ theory and the $+_{M,N}$ theory where the holomorphic functions $\cA_{\pm}$ are given by
\begin{align}
	 \cA_\pm^{T_N}&=\frac{3N}{8\pi} \left[\pm \ln(w-1)+i\ln(2w)+(\mp 1-i)\ln(w+1)\right]~,
	 \nonumber\\
	  \cA_\pm^{+_{N,M}}&=\frac{3}{8\pi}\left[iN\left(\ln(2w-1)-\ln(w-1)\right)\pm M \left(\ln (3w-2)-\ln w\right)\right]~,
\end{align}
The $T_N$ solution has three poles located at $w=\pm 1,0$ whereas the $+_{M,N}$ has four poles located at $w= 0,{1\over 2},{2\over 3},1$.
The relevant brane webs for the two theories are given by the junction of $N$ D5, $N$ NS5 and $N$ (1,1) 5-branes for the $T_N$ theory and the intersection of $N$ D5-branes and $M$ NS5-branes for the $+_{M,N}$  theory.

 \begin{figure}[h]
  \centering
  \includegraphics[width=50mm]{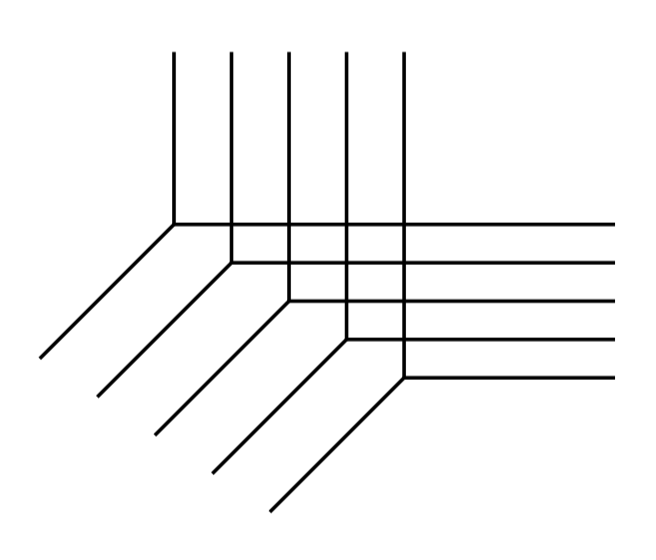}
    \includegraphics[width=50mm]{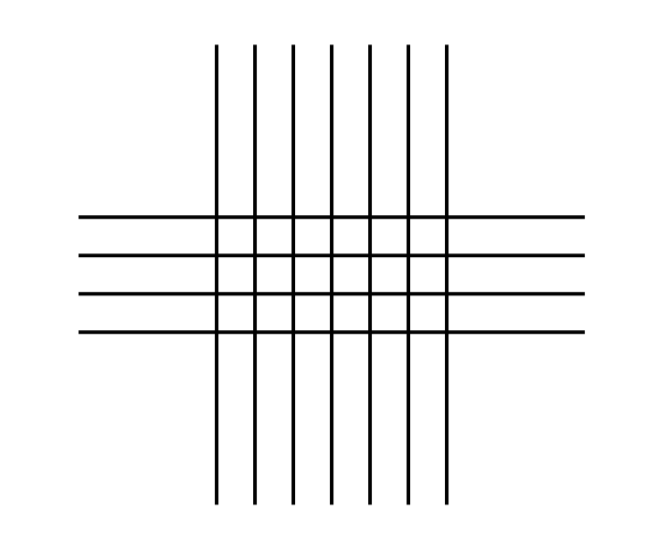}
  \caption{Left: brane web for the   $T_N$ theory. Right: brane web for the $+_{M,N}$ theory. }
\end{figure}
The quiver theories  which at their conformal fixed points realize the SCFTs are given by long  linear quivers with $SU(n)$ gauge nodes   and bi-fundamental matter connecting them and  a fundamental matter node at the end. 
For the $T_N$ theories we have  a linear quiver with $SU(k)$ nodes with increasing $k$ along the quiver and fundamental representations attached to the end of the quiver
\begin{align}
T_N:& \quad [2]-(2)-(3)-\cdots -(N-1)-[N]
\end{align}
For the $+_{MN}$ theory we have $M-1$ $SU(N)$ nodes with matter in the  fundamental representation of the end nodes  attached at the end of the quiver.
\begin{align}
+_{MN}: & \quad    [N]-(N)-(N)- \cdots -(N) -[N]   
\end{align}
For the two example cases the function $\cG$ defined in (\ref{eq:kappa2-G})   takes the  following compact form
\begin{align}
	\cG_{T_N} &=\frac{9}{8\pi^2}N^2D\left(\frac{2w}{w+1}\right)~,\nonumber\\
	\cG_{+_{N,M}} &=\frac{9}{8\pi^2}NM\left[D\left(\frac{3w-2}{w}\right)+D\left(\frac{w}{2-3w}\right)\right]
\end{align}
where $D$ is the Bloch-Wigner function given by
\begin{align}\label{eq:D-def}
	D(u)&=\Im\left[Li_2(u)+\ln(1-u)\ln |u|\right]~.\end{align}
	where $Li_2$ is the dilogarithm  function.

\section{Penrose limit and the plane wave background}
\label{sec3}

The idea behind taking a Penrose limit of $AdS_5\times S^5$ solution of type IIB  is to consider the trajectory of a particle sitting at the center of $AdS_5$ and moving very fast along a great circle on $S^5$,  such that in the limit the trajectory becomes a null geodesic. Zooming in on the vicinity of the null geodesic produces a pp-wave solution which preserves 32 supercharges.

For the $AdS_6\times S^2$ solutions  presented in section \ref{sec2}, the geometry is more complicated since the $AdS_6$ and $S^2$ are warped over the two dimensional surface $\Sigma$ and in general a nice Penrose limit does not exist for geodesics through a generic point on $\Sigma$. Furthermore as discussed in appendix  \ref{appnull} considering a geodesic with a nontrivial dependence on $\Sigma$ looks daunting due to the complicated form of the geodesic equation.

Fortunately, we can find a special point on the Riemann surface $\Sigma$ by considering critical points of the function $\cG$ defined in (\ref{eq:kappa2-G}), i.e  $w_c\in \Sigma$ for which
\begin{align}
\partial_w \cG\mid_{w=w_c } = 0 
\end{align}

It was observed in \cite{Gutperle:2020rty} that in all known examples, $ \cG$ has a unique critical point on the Riemann surface $\Sigma$.  For our two examples the critical point on $w_c$ on $\Sigma$ is located at

\begin{align}
w_{c,T_N} ={i\over \sqrt{3}}, \quad \quad w_{c, +_{M,N}}= {3+i\over 5}
\end{align}

In \cite{Gutperle:2020rty}  it was shown that the critical point $w_c$ correspond to the location of BPS probe D3-branes that realize surface defects in the dual SCFT. As we shall see in the following, the existence of the critical point is essential for our construction of  Penrose limit in the  warped spacetime.

We can examine our solution near the critical point by expanding  $w=w_c+ \epsilon  \zeta  $. The most important expression, $\cG$, can be expanded in a power series in $\epsilon$ 
\begin{align}\label{expana}
\cG&= G_0+ G_2  \epsilon^2 |\zeta|^2+ o(\epsilon^3)
\end{align}
Using (\ref{eq:kappa2-G}) we find the following expansions for the other functions which appear in the metric factors

\begin{align}\label{expanb}
\kappa^2&= - G_2+ o(\epsilon )\nonumber \\
T^2&=1-{2\over 3} {G_2\over G_0} \epsilon^2  |\zeta|^2 + o(\epsilon^3)\nonumber\\
R&= -{G_2\over 6 G_0} |\zeta ^2| + o(\epsilon)
\end{align}
Using the expressions for the metric factors (\ref{eq:metric-functions}) it is straightforward to show that the metric factors of the $AdS_6$ and $S_2$ are extremized at $w=w_c$.

For the  $T_N$ and $+_{N,M}$ examples we find the following expressions for the expansion coefficients $G_0$ and $G_2$
\begin{align}
T_N:&\;\; G_0=Im(Li_2(e^{i\pi/3})), \quad  G_2=- {81 \sqrt{3}\over 128  \pi^2}N^2\nonumber \\
+_{N,M}:&\;\; G_0 = {9 C M N\over 4 \pi^2}, \quad \quad G_2 =-{225 M N\over 16 \pi^2}
\end{align}
where $C$ is Catalan's constant.
\subsection{Penrose limit}\label{penroselim}

Using the expansions (\ref{expana}) and (\ref{expanb})   the Einstein frame metric  (\ref{eqn:ansatzmet}) can be expanded  to second order which corresponds to the metric close to the critical point $w=w_c$.
\begin{align}\label{expanmet}
ds^2 &=\sqrt{6G_0} \Big\{ (1+  \epsilon^2 {G_2\over 3 G_0} |\zeta|^2) (- \cosh^2 \rho dt^2 + d\rho^2 +\sinh^2\rho ds_{S^4}^2) \nonumber \\
&+ {1\over 9} (1+ \epsilon^2 {G_2\over  G_0} |\zeta|^2) (d\theta^2+ \cos^2\theta d\phi^2)+ {2\over 3} \epsilon^2 {|G_2|\over G_0} |d\zeta|^2\Big\}+ o(\epsilon^3)
\end{align}
We can drop the expansion parameter $ \epsilon$ in the following since it can be absorbed into a redefinition of  the coordinate 
$\zeta$.
The null geodesic which gives the proper Penrose limit is defined by
\begin{align}
\zeta =0,\quad  \rho=0, \quad  \theta=0, \quad  t-{\phi\over 3}=const
\end{align}
Note that $\zeta=0$ implies that the null geodesic is localized at the critical point $w=w_c$ on  the Riemann surface.
We
express the metric in terms of rescaled and relabelled coordinates
\begin{align}\label{penrosecv}
&t= x^++  {x^-\over (6G_0)^{1\over 2} }, \quad \quad  \phi=3 x^+-  {3x^-\over (6G_0)^{1\over 2} } \nonumber\\
& \rho = {1\over (6G_0)^{1\over 4} } r  , \quad \quad \theta = {3 \over  (6G_0)^{1\over 4} } x^6 , \quad \quad \zeta = \epsilon  \left( {3G_0\over 2 |G_2|}\right)^{1\over 2} {1\over  (6G_0)^{1\over 4} } (x^7+ i x^8)
\end{align}

The Penrose limit is obtained by taking $G_0\to \infty$ , which corresponds to taking $N\to \infty$ (for the $T_N$ case) and $MN\to \infty$ (for the $+_{N,M}$ example) and keeping the finite terms in the metric (\ref{expanmet}). All expressions involving the ratio $G_2/G_0$ are finite in this limit.
 Note  that the curvature radius of the $AdS_6$ and $S^2$ factors are controlled by $G_0$ and hence the limit of large $G_0$ is precisely the one taken for a Penrose limit in $AdS_p\times S^q$ spacetimes, i.e. zooming in on a region close to the null geodesic, which is much smaller than the curvature radius of the spacetime. 
 
 The resulting metric is given by
\begin{align}\label{planewavemetric}
ds^2 = -4 dx^+ dx^- + (dx^+)^2 \Big( - r^2 - 9 x_6^2 - x_7^2 -x_8^2\Big)  + dr^2 + r^2   ds_{S^4}^2 + dx_6^2+ dx_7^2+dx_8^2
\end{align}
where $r^2= x_1^2+x_2^3+ x_3^2+x_4^2+x_5^2$. The only difference between this metric and the one of the Penrose limit of $AdS_5\times S^5$ that the coordinate $x_6$ is singled out and has a different normalization in the $(dx^+)^2$ term in the metric, which in the world sheet action  leads to a different mass term for the field associated with $x_6$.

The  energy  in global AdS is associated with $i \partial_t$ whereas $-i \partial_\phi$ is related to a $U(1)$ generator of the $SU(2)_R$ symmetry, which is realized as the group of isometries of the $S^2$. Hence using the argument of  \cite{Berenstein:2002jq}  for the Penrose limit (\ref{penrosecv}) we can identify the light cone momentum with the conformal dimension and the R-charge of a dual operator via
\begin{align}\label{bpsop}
2p^-&= i \partial_+ = i (\partial_t + 3 \partial_\phi) =\Delta- 3 J \nonumber \\
2p^+& = i\partial_- = {i \over R^2} (\partial_t -3 \partial_\phi)= {1\over   R^2}(\Delta+ 3 J)
\end{align}
where we have identified the AdS radius with $R^2=\sqrt{6 G_0}$. Since we take $R$ to be very large in the Penrose limit excitations with finite $p^\pm$ correspond to states with large energy and angular momentum which on the CFT side correspond to operators with large conformal dimension $\Delta$ and R-charge $J$, which are close to the BPS bound $\Delta=3J$ \cite{Minwalla:1997ka,Cordova:2016emh}.

 \subsection{Other supergravity fields}

In the Penrose limit, the complex scalar field $B$ of type IIB supergravity  becomes a constant
\begin{align}
B&= -{\partial_w \cA_+\over \partial_w \cA_-}\mid_{w=w_c} + o(\epsilon)
\end{align}
Note that the standard RR axion $\chi$ and the dilaton $\phi$ is related to the complex scalar $B$ via
\begin{align}
B={1+ i\tau\over 1-i \tau}, \quad \quad \tau= \chi+ i e^{-\Phi}
\end{align}
Consequently, the axion dilaton has the following expression in terms of the holomorphic functions at the critical point $w_c$
\begin{align}
\tau = i {\partial_w \cA_-+\partial_w \cA_+ \over \partial_w \cA_-- \partial_w \cA_+}\mid_{w=w_c}
\end{align}
For the two examples we  consider in this paper  one finds\footnote{In \cite{DHoker:2016ujz,DHoker:2017mds}  different normalization for the dilaton is used, we have translated the expression to the commonly used one.} 
\begin{align}
T_N: & \;\; \chi={1\over 2}, \;\; e^{-\phi}={\sqrt{3}\over 2}, & +_{N,M}: \;\; \chi=0,\;\; e^{-\phi} = {N\over M}
\end{align}
The complex rank two antisymmetric tensor potential $C^{(2)}$ is related to the NS-NS and R-R two form potentials by
\begin{align}
C^{(2)}&=B^{NSNS}+ i C^{RR} \nonumber\\
&= \Big( c_0  -i  {8\over 9}\Big( \partial_w \cA_+\mid_{w=w_c}  \epsilon\zeta +\partial_{\bar w } \bar \cA_-\mid_{w=w_c} \epsilon \bar \zeta \Big) + o(\epsilon^2) \Big)   \cos \theta d\theta \wedge d\phi 
\end{align}
  Note that the constant term $c_0$  is not important since in the bosonic part of the worldsheet action, a constant  NS-NS anti-symmetric tensor  potential will be a total derivative. In the fermionic worldsheet action only the field strength, which does not depend on $c_0$, appears.   In the Penrose limit the  finite part which survives  (after using the rescaled variables)
\begin{align}
C^{(2)}= -4i {1\over |G_2|^{1\over 2} } \Big( \partial_w \cA_+\mid_{w=w_c} (x^7+ i x^8)  +\partial_{\bar w } \bar \cA_-\mid_{w=w_c} (x^7-i x^8) \Big)dx^6\wedge dx^+
\end{align}
Note that in the scaling limit which produces the Penrose limit discussed below (\ref{penrosecv})  the $\partial A_\pm $ and $|G_2|$ scale in the same way and hence  the term above is the expression  which  survives the Penrose limit.

The NS-NS 3 form field strength $H^3_{NS}$ and the RR field strength $F^3_{RR}$ are given by

\begin{align}
H^3_{NS} =Re\Big( dC_2\Big) ,\quad \quad F^3_{RR} =Im\Big( dC_2\Big) ,
\end{align}
For the examples we get
\begin{align}
T_N:&\quad   H^3_{NS} = -{4 \sqrt{2} \over 3^{1\over4}  }dx^8\wedge dx^6 \wedge dx^+ ,\nonumber\\
& \quad  F^3_{RR}= -{2\sqrt{2} \over 3^{3\over4} } \Big( 3 dx^7+ \sqrt{3} dx^8) \wedge d x^6 \wedge dx^+
\end{align}
and
\begin{align}
+_{N,M}:& \quad H^3_{NS} = - {4  \over 5} \sqrt{M\over N} ( 4 dx^7+ 3 dx^8) \wedge d x^6 \wedge dx^+ ,\nonumber\\
& \quad  F^3_{RR}= {4  \over 5} \sqrt{N\over M} ( -3 dx^7+4dx^8) \wedge d x^6 \wedge dx^+
\end{align}

It is important to mention that the supergravity solutions of section \ref{sec2}  are given in the Einstein frame. In order to  quantize the Green-Schwarz string in this background we have to transform to the string frame
\begin{align}\label{string-frame}
    G_{\mu\nu}^{string} = e^{\phi /2} G_{\mu\nu}^{Einstein}
\end{align}
Note that the particular combinations of the anti-symmetric tensor field strength which appear  in the string frame supersymmetry transformations as well as the  Green-Schwarz action are $H^{(3)}_{NS}$ and $e^{\phi}(F^{(3)}_R-\chi H^{(3)}_{NS})$.

In the Penrose limit, the transformation to the Einstein frame is just a constant rescaling of the metric. However, it is convenient for the string frame metric in to be the canonical plane wave form (\ref{planewavemetric}). This can be  accomplished by a rescaling of all coordinates (except $x^+$) by a factor of $e^{-\phi/4}$. This has the effect of introducing a factor of $e^{-\phi /2}$ in each of the antisymmetric tensor fields. The end result is  the following combination of fields which appears in the string worldsheet:

\begin{align}\label{Tnfields}
T_N:&\quad   H^3_{NS, string} = -4  dx^8\wedge d x^6 \wedge dx^+ , \nonumber\\  &e^{\phi}( F^3_{RR, string}-\chi H^3_{NS, string})= -4 dx^7  \wedge dx^6 \wedge dx^+  \\
+_{N,M}:& \quad  H^3_{NS, string} = - {4  \over 5}  ( 4 dx^7+ 3 dx^8) \wedge dx^6 \wedge dx^+ , \nonumber\\ \label{+mnfields}
& e^{\phi}( F^3_{RR, string}-\chi H^3_{NS, string})= {4  \over 5} ( -3 dx^7+4dx^8) \wedge d x^6 \wedge dx^+
\end{align}
We will drop "string" from the subscripts moving forward since we will always be working in the string frame and there will be no confusion.

Note one interesting property  of  specific solutions (\ref{Tnfields}) and (\ref{+mnfields}): The orientation of the forms in the $7,8$ directions can be parameterized as $n_7 dx^7+ n_8 dx^8$, for  both $T_N$ and $+_{MN}$   examples the $n_a$ associated with the $H^{(3)}_{NS}$ and $e^{\phi}(F^{(3)}_R-\chi H^{(3)}_{NS})$ tensors are orthogonal to each other and square to four. 
This means that we could rotate $x^{7,8}$ so that the fields which appear in the worldsheet action take exactly the same form in both $T_N$ and $+_{MN}$ theories and therefore give the same string spectrum. In fact, this statement also holds for all other solutions \cite{Bergman:2018hin, Gutperle:2017tjo} that we have checked. We conjecture that all global solutions (\ref{eqn:ansatzmet}) - (\ref{bc2sol})  share this property and in the quantization the follows, take our fields to have the form of the $T_N$ example (\ref{Tnfields}).

\section{Light cone Green-Schwarz string spectrum}
\label{sec4}

In this section we quantize the Green-Schwarz string  the pp-wave spacetime obtained in the previous section. As shown in \cite{Metsaev:2002re, Metsaev:2001bj,Mizoguchi:2002qy, Russo:2002rq} the fermionic part of the Green-Schwarz string becomes quadratic in fermionic fields for the pp-wave spacetimes in the light cone gauge, which makes the free string spectrum exactly solvable.
\subsection{Bosonic Spectrum}

We will start by examining the bosonic spectrum for the Green-Schwarz string in the plane wave background with NS-NS flux. The general action is given by 

\begin{align}
    S_{b}=-\frac{1}{4 \pi \alpha^{\prime}} \int d^{2} \sigma \sqrt{h}\left(h^{a b} \partial_{a} X^{\mu} \partial_b X^{\nu} G_{\mu \nu}+\varepsilon^{a b} \partial_{a} X^{\mu} \partial_{b} X^{\nu} B_{\mu \nu}\right)
\end{align}

As in the $AdS_5\times S^5$ case, this action simplifies considerably after light cone gauge-fixing, leaving us with a free theory which can be easily quantized following \cite{Polchinski:1998rq}. We set the target space coordinate $x^+ = \tau$ and worldsheet metric to $h^{\sigma\tau} = 0 $, $h^{\sigma\sigma} = - h^{\tau\tau} = 1$ and after plugging in the background (\ref{planewavemetric}), (\ref{Tnfields}), we are left with the following action for the eight remaining transverse scalar fields

\begin{align}
    S_b=\frac{1}{4 \pi \alpha^{\prime}} \int_{0}^{2 \pi \alpha^{\prime} p^{+}} d^2 \sigma\left(\sum_{I=1}^{8}\left(\left(\partial_{\tau} X^{I}\right)^{2}-\left(\partial_{\sigma} X^{I}\right)^{2}-m_{I}^{2}\left(X^{I}\right)^{2}\right)+4 X^{6} \partial_{\sigma} X^{7}-4 X^{7} \partial_{\sigma} X^{6}\right)
\end{align}
where $m_I^2 = 1$ for all $I\neq 6$ in which case $m_6^2 = 9$.  The equations of motion deduced from this action are given by
\begin{align}
\partial_{\tau}^{2} X^{I}-\partial_{\sigma}^{2} X^{I}+X^{I} &=0  \quad I = 1, ... , 5,8\label{x18eom}
\\ \label{x6eom}
\partial_{\tau}^{2} X^6-\partial_{\sigma}^{2} X^6+9X^6- 4\partial_{\sigma}X^7 &=0 \\ \label{x7eom}
\partial_{\tau}^{2} X^7-\partial_{\sigma}^{2} X^7+X^7 + 4\partial_{\sigma}X^6 &=0 
\end{align}
Together with the periodicity condition $X^I(\sigma + 2\pi\alpha 'p^+ , \tau) = X^I (\sigma,\tau)$, the first set of equations (\ref{x18eom}) lead to the following familiar solutions for $I = 1,...,5,8$

\begin{align}\label{oscillatorsolution}
    x^I (\sigma,\tau) = x_{0}^{I} \cos  \tau+\frac{p_{0}^{I}}{ p^{+}} \sin  \tau+\sqrt{\frac{\alpha^{\prime}}{2}} \sum_{n=1}^{\infty} \frac{1}{\sqrt{\omega^I_{n}}}[& \alpha_{n}^{I} e^{-\frac{i}{\alpha^{\prime} p^{+}}\left(\omega^I_{n} \tau+n \sigma\right)}+\tilde{\alpha}_{n}^{I} e^{-\frac{i}{\alpha^{\prime} p^{+}}\left(\omega^I_{n} \tau-n \sigma\right)}+ \nonumber\\
&\alpha_{n}^{I \dagger} e^{\frac{i}{\alpha^{\prime} p^{+}}\left(\omega^I_{n} \tau+n \sigma\right)}+\tilde{\alpha}_{n}^{I \dagger} e^{\frac{i}{\alpha^{\prime} p^{+}}\left(\omega^I_{n} \tau-n \sigma\right)}]
\end{align}
where the frequencies will determine the energy of the string excitations
\begin{align}
    \omega_{n}^{I}=\sqrt{n^2 + (\alpha' p^+)^2} 
\end{align}
To solve the remaining two equations of motion (\ref{x6eom}) and  (\ref{x7eom}) for $I=6,7$  we use an ansatz of the following form
\begin{align}\label{ansatzoscillators}
    x^I(\sigma,\tau) =  \sqrt{{\alpha' \over 2}} A_0(\tau) + \sqrt{{\alpha' \over 2}} \sum_{n>0} [A_n^I(\tau, \sigma) + \tilde A_n^I(\tau, \sigma) ]
\end{align}
where $A^I_n$ and $\tilde A^I_n$  for $I=6,7$ can be expressed in terms of eigenmodes 
\begin{align}\label{aneigen}
A_n^I (\tau, \sigma) &= \sum_{J=6,7} {1 \over \sqrt{\omega_n^J}} \Big\{ (V_n)^I_{J}\alpha_n^{J}e^{-{i\over \alpha'p^+}(\omega_n^J\tau+n\sigma)}+ (\bar V_{n})^I_{J}(\alpha^\dagger)_n^{J}e^{{i\over \alpha'p^+}(\omega_n^J\tau+n\sigma)}\Big\}  \nonumber\\
\tilde A_n^I (\tau, \sigma) &= \sum_{J=6,7} {1 \over \sqrt{\omega_n^J}} \Big\{ (V_{-n} )^I_{J}\alpha_n^{J}e^{-{i\over \alpha'p^+}(\omega_n^J\tau-n\sigma)}+ (\bar V_{-n})^I_{J}(\alpha^\dagger)_n^{J}e^{{i\over \alpha'p^+}(\omega_n^J\tau-n\sigma)}\Big\}  
\end{align}
For the zero mode $A_0$ there only one eigenfunction which can be obtained from $A_n^I$  in  (\ref{aneigen}) by setting $n=0$. The eigenfrequencies for $J=6,7$
 \begin{align}
\omega_{n}^6&= \sqrt{ n^2 +(\alpha'p^+)^2}+2\alpha' p^+ \nonumber\\
\omega_{n}^7 &= \left|\sqrt{ n^2 +(\alpha'p^+)^2}-2\alpha' p^+\right|
\end{align}
and $(V_n)^I_{\;J}, J=6,7 $ are the orthonormal set of  vectors which  satisfy

\begin{align}
\left(
\begin{array}{cc}
  {-(\omega^J_n)^2 + n^2\over (\alpha' p^+)^2}+9&   -  {4 i  n\over \alpha'p^+}    \\
 + {4 i  n\over \alpha'p^+} &         {-(\omega^J_n)^2 + n^2\over (\alpha' p^+)^2}+1
\end{array}
\right) \left(
\begin{array}{c}
(V_n)^6_{\;J }\\
(V_n)^7_{\;J}
\end{array}
\right)=0, \quad \quad J=6,7
\end{align}
  
With this mode equation the the canonical quantization conditions  
\begin{align}
[x^I(\sigma, \tau), p^J(\sigma ' ,\tau)] = i\delta ^{IJ}\delta(\sigma-\sigma')
\end{align}
where $p^I = {1\over 2\pi\alpha ' }\partial_{\tau} x^I$, yield the commutation relation for the creation and annihilation operators
\begin{align}
    [\alpha_n^I , \alpha_m^{J \dagger}] = [\Tilde{\alpha}_n^I , \Tilde{\alpha}_m^{J \dagger}] = \delta^{IJ} \delta_{mn}
    \end{align}
where now we allow the indices $I,J$ to run through all eight transverse oscillators $1,2, \cdots, 8$. 

The light-cone Hamiltonian can be expressed in terms of the creation and annihilation operators (\ref{oscillatorsolution}) and (\ref{ansatzoscillators})
\begin{align}\label{hambos}
    H_{l . c .}^{\text {b }}= \frac{1}{\alpha^{\prime} p^{+}}&\sum_{I=0}^8\left[ \omega^I_0\alpha_0^{I\dagger}\alpha_0^I + \sum_{n=0}^{\infty} \omega_{n}^I\left(\alpha_{n}^{I \dagger} \alpha_{n}^{I}+\tilde{\alpha}_{n}^{I \dagger} \tilde{\alpha}_{n}^{I}\right) \right] + { \nu_{\rm bos}}\end{align}
   Here $ \nu_{\rm bos}$ denotes a normal ordering constant for the bosonic creation and  annihilation operators.
The  zero mode modes $x^I_0, p^I _0, I=1,2,\cdots,5,8$ have been expressed    in terms of creation and  annihilation operators  
\begin{align}
    \tilde{\alpha}_{0}^{I}=\frac{1}{\sqrt{2  p^{+}}} p_{0}^{I}-i \sqrt{\frac{ p^{+}}{2}} x_{0}^{I}
\end{align}
to maintain consistent normalization. It will be convenient to write $\nu_{\rm bos}$ in terms of the oscillator frequencies:
\begin{align}\label{zeropoint}
  \nu_{\rm bos}={1\over 2 \alpha'p^+}\sum_{I=1}^8\big( \omega^I_0+ 2 \sum_n  \omega_n^I\big)
  \end{align}

\subsection{Fermionic Spectrum}
We can now turn our attention to the fermionic part of the Green-Schwarz action for type IIB strings, the term quadratic in the world sheet fermions is given by  (see appendix \ref{appa} for details on the notation).

\begin{align}
    S_{f}^{(2)}=\frac{i}{4 \pi \alpha^{\prime}} \int d^{2} \sigma \sqrt{h}\left(h^{i j} \delta_{a b}-\varepsilon^{i j}\left(\sigma_{3}\right)_{a b}\right) \partial_{i} x^{M} \bar{\theta}^{a} \Gamma_{M}\left(D_{j}\right)_{c}^{b} \theta^{c}
\end{align}

This Green-Schwarz action possesses a  $\kappa$ symmetry which is  required to obtain spacetime supersymmetry for the on-shell string modes \cite{Metsaev:2001bj}. To obtain these physical fermionic modes, we can gauge fix (as in flat space) by choosing 
\begin{align}\label{lcferm}
    \Gamma^+ \theta^a = 0 
\end{align}
where $\Gamma^{\pm} = {1\over\sqrt{2}}(\Gamma^0 \pm \Gamma^9)$.  After enforcing the light cone gauge condition $x^+ = \tau$ and worldsheet metric to $h^{\sigma\tau} = 0 $, $h^{\sigma\sigma} = -h^{\tau\tau} = 1$, the fermionic part of the Green-Schwarz action reduces to the quadratic one in the plane wave background \cite{Metsaev:2001bj,Metsaev:2002re,Mizoguchi:2002qy}  and  takes the following form 
\begin{align}
    S_f=\frac{i}{4 \pi \alpha^{\prime}} \int d^{2} \sigma\left(-\left(\theta^{a}\right)^{T} \Gamma^{+} \Gamma_{+}\left(D_{\tau}\right)_{c}^{a} \theta^{c}-\left(\theta^{a}\right)^{T} \Gamma^{+} \Gamma_{+} \sigma_{a b}^{(3)}\left(D_{\sigma}\right)_{c}^{b} \theta^{c}\right)
\end{align}
after inserting the  antisymmetric tensor fields (\ref{Tnfields}) for the $T_N$ theory, the action becomes
\begin{align}\label{TNaction}
    S_f&=- {i\over 4 \pi \alpha'} \int  d^2 \sigma (\theta^a)^T \Gamma^+ \Gamma_+ \Big(  \delta_{ab} \partial_\tau   +  (\sigma_3)_{ab} \partial_\sigma  -   {1\over 4} H_{+ 67}\Gamma^{67} (\sigma_3)_{ab} +{1\over 4} \tilde F_{+ 68}  \Gamma^{68}(\sigma_1)_{ab}\Big)\theta^b\nonumber \\
&= - {i\over 4 \pi \alpha'} \int  d^2 \sigma (\theta^a)^T \Gamma^+ \Gamma_+ \Big(  \delta_{ab} \partial_\tau   +  (\sigma_3)_{ab} \partial_\sigma  +   \Gamma^{67} (\sigma_3)_{ab} - \Gamma^{68}\ (\sigma_1)_{ab}\Big)\theta^b
\end{align}
where we have used the fact that  $\Gamma^+\Gamma_+=1+ \Gamma^{09}$ and hence 
$(\Gamma^+\Gamma_+)^2 = 2\Gamma^+\Gamma_+$. 
The chirality condition $\Gamma^{11} \theta^a=\theta^a$ together with the light cone gauge condition (\ref{lcferm}) projects a general thirty-two component spinor onto an eight dimensional subspace. It is possible to find a representation of the Gamma matrices in this subspace and combine them with the $\sigma$  matrices encoding the two chiral spinors  such that the action can be written as
\begin{align}\label{sfermb}
 S_f&=- {i\over 2 \pi \alpha'} \int  d^2 \sigma \theta^T 1_4\otimes \Big(1_2\otimes 1_2\; \partial_\tau+  1_2\otimes \sigma_3\; \partial_\sigma+\sigma_3 \otimes \sigma_3 - \sigma_1 \otimes \sigma_1 \Big)\theta
\end{align}

Due to the common $1_4$ factor in all the terms it follows that there is a fourfold degeneracy and 
equation of motion for each  degenerate spinor component is given by

\begin{align}\label{eqofmf}
\Big(  1_2\otimes 1_2\; \partial_\tau + 1_2\otimes \sigma_3\; \partial_\sigma + i  \sigma_3 \otimes \sigma_3 -  i \sigma_1 \otimes \sigma_1 \Big)\theta^\alpha =0
\end{align}
where $\alpha = 1,...,4$ labels four degenerate spinor components acting on the first $1_4$ factor in the tensor product.
We can proceed just as in the bosonic case with an ansatz of the form
\begin{align}
    \theta^{\alpha}(\sigma,\tau) = {1\over \sqrt{2 p^+} }   \sum_{n=0}^\infty  \Big(\theta_n^{\alpha,+} (\tau, \sigma)   +\theta_n^{\alpha,-} (\tau, \sigma)  \Big) 
    \end{align}
    Where
   
    \begin{align} \label{eigeferma}
    \theta_n^{\alpha,\pm }  &= V^\pm_n e^{-{i\over \alpha'p^+}(\omega_n^\pm \tau+n\sigma)}\beta_n^{\alpha,\pm} + V^\pm_{-n} e^{{i\over \alpha'p^+}(\omega_n^\pm \tau+n\sigma)}(\beta_n^{\alpha,\pm})^\dagger\nonumber \\
    &\quad  + V^\pm_{-n} e^{-{i\over \alpha'p^+}(\omega_n^\pm \tau-n\sigma)}\tilde \beta_n^{\alpha,\pm} +V^\pm_{n} e^{{i\over \alpha'p^+}(\omega_n^\pm \tau-n\sigma)}\tilde ( \beta_n^{\alpha,\pm})^\dagger 
    \end{align} 
    the positive eigenfrequencies are given by
    \begin{align}\label{fermionfrequencies}
    \omega_n^{+} = \sqrt{(\alpha' p^+)^2 + n^2}+  \alpha' p^+ \nonumber\\ 
    \omega_n^{-} =  \sqrt{(\alpha' p^+)^2 + n^2} - \alpha' p^+
\end{align}
The vectors $V^\pm_n$ satisfy the following equation
\begin{align}
\Big(  -   \omega_n^\pm\; 1_2\otimes 1_2\;-  n  \; 1_2\otimes \sigma_3  +  \sigma_3 \otimes \sigma_3 -  \sigma_1 \otimes \sigma_1 \Big)V_n^\pm =0
\end{align}
 ensuring that the mode expansion of $\theta$  (\ref{eigeferma}) satisfies the equation of motion (\ref{eqofmf}).
For each $n>0$   the  four vectors $V^\pm_n, V^\pm_{-n}$ can be chosen to form a orthonormal basis. 
The canonical anti-commutation relations following from the action (\ref{sfermb})
\begin{align}
\{ \theta^T (\tau, \sigma), \theta(\tau, \sigma')\} = i 2\pi \alpha' \delta(\sigma-\sigma') 1_{16\times16}
\end{align}
imply the following algebra for the fermionic creation and annihilation operators
\begin{align}
    \{\beta_n^{\alpha,\pm },(\beta_m^{\gamma, \pm })^{\dagger}\} =  \{\tilde \beta_n^{\alpha,\pm },(\tilde \beta_m^{\gamma, \pm })^{\dagger}\}=  \delta^{\alpha\gamma}\delta_{m,n}
\end{align}
with all other anti commutators vanishing.

Interestingly, we can see from (\ref{fermionfrequencies}) that our spectrum contains eight fermionic zero modes with $\omega^-_n= 0$ when $n=0$. The zero modes are associated with $\theta$  in the original fermionic action (\ref{TNaction})
 for which
\begin{align}
    \left(\Gamma^{67} \sigma^{(3)} -\Gamma^{68} \sigma^{(1)} \right) \theta=\Gamma^{67} \sigma^{(3)}\left(1-i \Gamma^{78} \sigma^{2}\right) \theta=0
\end{align}
where we have dropped the index $a$ for notational simplicity.
Notice that $\left(1-i \Gamma^{78} \sigma^{2}\right)$ is a projector and hence has an eight dimensional eigenspace with eigenvalue zero, matching precisely with the eight zero modes found above. 

Lastly, we can write the fermionic light cone Hamiltonian 
\begin{align}\label{hamferm}
    H_{l . c .}^{f}=\frac{1}{\alpha^{\prime} p^{+}}&\sum_{\alpha}\left[ \omega_0^{ +}(\beta_{0}^{\alpha,+})^{\dagger} \beta_{0}^{\alpha,+} + \sum_{n=0}^{\infty}\sum_{\gamma=\pm}  \omega^{\alpha,\gamma}_{n}\left((\beta_{n}^{\alpha,\gamma})^{\dagger} \beta_{n}^{\alpha+} +(\tilde \beta_{n}^{\alpha,\gamma})^{\dagger} \tilde \beta_{n}^{\alpha,\gamma} \right) \right] +\nu_f
\end{align}
Where $\nu_f$ is the  constant obtained by  normal ordering the fermionic creation and anilhilation operators.

\begin{align}\label{noferm}
\nu_f&= -{1\over 2 \alpha' p^+} \Big( 4\omega_0^+ + 8\sum_{n>0} (\omega_n^+ +\omega_n^-)\Big) \nonumber  
\end{align}

\subsection{Light cone spectrum}\label{lcspec}

The Hamiltonian of the light cone string is the sum  of the bosonic (\ref{hambos}) and fermionic (\ref{hamferm}) parts
\begin{align}
H_{l . c .}&=\frac{1}{\alpha^{\prime} p^{+}} \left( \sum_{I=0}^8\left[ \omega^I_0\alpha_0^{I\dagger}\alpha_0^I + \sum_{n=1}^{\infty} \omega_{n}^I\left(\alpha_{n}^{I \dagger} \alpha_{n}^{I}+\tilde{\alpha}_{n}^{I \dagger} \tilde{\alpha}_{n}^{I}\right) \right]  \right. \nonumber\\
&\left. \quad  \quad +\sum_{\alpha}\left[ \omega_0^{ +}(\beta_{0}^{\alpha,+})^{\dagger} \beta_{0}^{\alpha,+} + \sum_{n=1}^{\infty}\sum_{\gamma=\pm}  \omega^{\alpha,\gamma}_{n}\left((\beta_{n}^{\alpha,\gamma})^{\dagger} \beta_{n}^{\alpha+} +(\tilde \beta_{n}^{\alpha,\gamma})^{\dagger} \tilde \beta_{n}^{\alpha,\gamma} \right) \right] \right) +\nu
\end{align}
The normal ordering  constant is the sum of the bosonic  (\ref{zeropoint}) and fermionic  (\ref{noferm}) contribution and  the bosonic and fermionic normal ordering contributions  cancel up to a finite number of terms
\begin{align}
\nu&= \nu_b+\nu_f \nonumber \\
&={1\over \alpha'p^+} \Big( 2 \alpha'p^+ + 4 \sum_{0<n<\sqrt{3} \alpha'p^+} \big( 2 \alpha'p^+ -\sqrt{n^2+ (\alpha'p^+)^2}\big)\Big)
\end{align}
It is interesting to note that in the limit of vanishing $\alpha'p^+$ the finite normal ordering constant goes to zero, in agreement with the zero normal ordering constant for the light cone string in flat space.  On the other hand for large $\alpha'p^+$, a large but finite number of modes contribute to the normal ordering constant\footnote{The fact that the normal ordering constant is not divergent can be understood from the one loop ultraviolet finiteness of the Green-Schwarz string.}.

In the light cone gauge the level matching constraint is obtained by considering the variation of the action  with respect to $h^{\tau\sigma}$ in terms of the bosonic and fermionic modes one finds the constraint
\begin{align}
 \sum_{n=1}^{\infty} \omega_{n}^I\left(\alpha_{n}^{I \dagger} \alpha_{n}^{I}-\tilde{\alpha}_{n}^{I \dagger} \tilde{\alpha}_{n}^{I}\right)+\sum_{n=1}^{\infty}\sum_{\gamma=\pm}  \omega^{\alpha,\gamma}_{n}\left((\beta_{n}^{\alpha,\gamma})^{\dagger} \beta_{n}^{\alpha+} -(\tilde \beta_{n}^{\alpha,\gamma})^{\dagger} \tilde \beta_{n}^{\alpha,\gamma} \right) =0
\end{align}
The spectrum of the light cone string has several features which distinguish it from the Penrose limit of $AdS_5\times S^5$  discussed in \cite{Berenstein:2002jq}. First due to the absence of world sheet supersymmetry the bosonic and fermionic worldsheet energies $\omega^I_n$ and $\omega^{\alpha, \pm}_{n}$ are not the same. As discussed in  appendix \ref{apa1} this is  a consequence of the fact that there are no  "supernumerary"  \cite{Cvetic:2002nh} supersymmetries in the pp-wave background we consider.  A second difference is the presence of fermionic  zero modes associated with the $\beta_0^{\alpha,-}, (\beta_0^{\alpha,-})^\dagger$ modes. These modes imply that even though the bosonic and fermionic oscillators have different energies that there are an equal number of bosonic and fermionic states at each energy level due to the 8 bosonic and 8 fermionic states created by $(\beta_0^{\alpha,-})^\dagger, \;\alpha=1,2,3,4$. The third feature is the presence of a normal ordering constant and the dependence on $\alpha' p^+$. We discuss some possible reasons for the different properties of the lightcone string action in the next section.  Presently, we have not been able to find an  identification of the string spectrum with BMN like operators on the field theory side.

\section{Discussion}
\label{sec5}
In this paper we have constructed a Penrose limit for  type IIB  warped $AdS_6\times S^2$ solutions. We have constructed the limit by zooming in on a null geodesic at the center of the $AdS_6$ and along a great circle of the $S^2$. This construction only works if the geodesic is localized at the critical point $w=w_c$  of the function $\cG$ on the two dimensional Riemann surface $\Sigma$. We constructed the pp-wave supergravity solution one obtains from taking the Penrose limit.   After appropriate coordinate transformation the resulting pp-wave background is universal for all the cases we have considered.  It would be interesting to consider a  more general class of examples, such as the solutions including seven branes \cite{DHoker:2017zwj} or O7-branes 
\cite{Uhlemann:2019lge} and confirm that the behavior of the pp-wave at the critical point is of the same form.
Another observation  is that all known regular solutions have a unique critical point $w_c$ on the Riemann surface. This statement has not been proven but checked in all  cases constructed in \cite{Fluder:2020pym}.  It is interesting that this critical point $w_c$ is also relevant for supersymmetric embedding of D3-branes in the $AdS_6\times S^2$ solutions which realizes BPS co-dimension two defects in the dual SCFT  \cite{Gutperle:2020rty}.

We presented the  quadratic bosonic and fermionic world sheet actions of the Green-Schwarz string in this background and calculated the spectrum of bosonic and fermionic excitations of the string in the light cone gauge. The spectrum has some interesting features which are different from other cases such as the Penrose limit of $AdS_5\times S^5$.  Namely that the frequencies associated with the bosonic and fermionic creation and annihilation operators do not coincide, which is a consequence of the absence of any "extra" supersymmetries which would be associated with linearly realized supersymmetries.  An additional new feature is the presence of fermionic zero modes, which makes sure that there are equal number of fermions and bosons at each level of the light cone hamiltonian.

One of the exciting features of the BMN correspondence was the identification of the light cone vacuum with a special class of BPS operators in the N=4 SYM theory, which can be viewed as a long spin chain. Furthermore, the action of creation operators on the string vacuum can be related to the insertion of  impurities into the spin chain and diagonalizing the Hamiltonian. It is an interesting question whether a similar identification can be found for the light cone vacuum and excited states of the Green-Schwarz  string in this plane wave background. This identification is more challenging for two reasons. While the Penrose limit  identifies the light cone string vacuum with BPS states which satisfy $\Delta =3J$, where $J$ is a $U(1)$ inside the $SU(2)_R$ symmetry of the SCFT and therefore related to protected BPS multiplet, these theories are realized as conformal fixed points of long quiver theories and hence strongly coupled.
For $d=4,\cN=4$ SYM the identification of these operators was possible in terms of the scalar fields of the SYM theory. For long quiver theories it is not clear how to construct these operators from the fundamental fields of the nodes of the quiver and the bi-fundamental matter.   A  class of such operators  corresponding to "stringy" long meson and baryon operators  was given in \cite{Bergman:2018hin} which  are constructed by taking   products of the bi-fundamental 
hyper multiplets, from one end of the quiver to the other\footnote{See \cite{Baume:2020ure} for a related construction in theories dual to six dimensional SCFTs.}. Since our null geodesic is at a fixed location on $\Sigma$ is is natural to speculate that the BMN like operator would be associated with a single node in the quiver unlike the string like operators in  \cite{Bergman:2018hin}  which stretch across the Riemann surface $\Sigma$.  It would be interesting to investigate what mechanism on the field theory side singles out a specific node in the quiver theory. One could also investigate more general null geodesics which are not at a fixed point on $\Sigma$ and consider the their Penrose limits. A very preliminary discussion of this issue  can be found in appendix \ref{appnull}.

In \cite{Uhlemann:2020bek,Uhlemann:2019ypp,Fluder:2018chf} sphere partition functions and expectation values of BPS Wilson line operators  for five dimensional SCFTs were calculated using localization techniques. For long quivers in the limit of large number of nodes and large gauge groups,  the eigenvalues for the nodes where replaced by a continuous distribution and the saddle point of the path integral was reduced to an analogue electrostatic problem.  It would be interesting to see whether these methods could be adapted to identify the BMN operators or whether the two dimensional electrostatic problem can be related to the light cone world sheet in some way and explain some of the curious properties of the light cone spectrum described in section \ref{lcspec}. 

We leave these interesting questions for future work.

\section*{Acknowledgements}
The work of M.G.~was  supported, in part,  by the National Science Foundation under grant PHY-19-14412.  We  are grateful to  the Mani L. Bhaumik Institute for Theoretical Physics for support. We are grateful to C. Uhlemann for useful discussions and comments on a draft of this paper.

\newpage

\appendix

\section{Supersymmetry of the plane wave background} \label{appa}

 The fermionic supersymmetry transformations  in the string frame can be expressed in terms of the following two operators  \cite{Grana:2002tu}, that are also used to construct the part of the Green-Schwarz string action which is quadratic in space time fermions. 
\begin{align}\label{doper}
D_{\; b}^a &= -{1\over 4\cdot 3!} H_{mnp}\Gamma^{m np} (\sigma^3)_{\;a}^{b} -{1\over 4 \cdot 3!} e^{\phi} (F_{mnp}- \chi H_{mnp})\Gamma^{mnp} (\sigma^1)_{\; a}^{b} \nonumber\\
(D_M)_{\; b}^a &= \partial_M +{1\over 4} \omega_{mn,M} \Gamma^{mn}\delta_a^b -{1\over 8} e^m_M H_{mnp}\Gamma^{np} (\sigma^3)_{\; a}^{b} + {1\over  8 \cdot 3!} e^{\phi} (F_{npq}- \chi H_{npq})\Gamma^{npq}  \Gamma_M (\sigma^1)_{\; a}^{b} 
\end{align}

Here $\sigma^a, a=1,3$ are the Pauli matrices which act on   two component  Majorna-Weyl spinors\footnote{The dictionary to go from the two component (with the matrices $\sigma $ to formulation of using  complex Weyl spinors spinor  is given by $\epsilon=\epsilon^1+i \epsilon^2$ and 
$\sigma^1 \epsilon\to i \epsilon^*, \quad i \sigma ^2 \to -i \epsilon , \quad \sigma^3 \epsilon\to \epsilon^*$.}. 
The supersymmetry transformations  in the two component formalism are given by 

\begin{align}\label{generalSUSYtransform}
\delta\lambda^a = D_{\; b}^a \epsilon^b, \quad \quad \delta \psi_M ^a= (D_M)_{\; b}^a \epsilon^b
\end{align}
The susy transformations for the gravitino in  the string frame are is different from the one in the Einstein frame given in \cite{Schwarz:1983qr}, this is because the transformation $g_{MN} \to e^{\phi/2}g_{MN}$ induces a mixing of the dilatino supersymmetry transformation with the gravitino supersymmetry transformation  \cite{Hassan:1999bv}. The supersymmetry transformations (\ref{generalSUSYtransform}) are given in the string frame.

The string world sheet action which is quadratic in fermions is given by

\begin{align}\label{Lrr}
S^{(2)}_{RR}={1\over 4 \pi \alpha'} \int  d^2 \sigma \sqrt{h} \Big(h^{ij}\delta_{ab}-\epsilon^{ij}(\sigma^3)_{ab}\Big)\partial_i x^M \bar \theta^a \Gamma_M ( D_j )^{b}_{\; c} \theta^c
\end{align}

where $\theta^a$, $a=1,2$ are the two ten dimensional  Weyl-Majorana spinor  world sheet fields with the same chirality as the supersymmetry transformation parameters $\epsilon^a$ in  (\ref{generalSUSYtransform}) . The operator $(D_j)^b_c$ is the pullback of the covariant derivative to the world sheet   is related to (\ref{doper}) by
\begin{align}\label{covariantderivative}
( D_j )^{b}_{\; c}&= \partial_j  \delta^a_{\; b} + \partial_j x^M\Big( {1\over 4} \omega_{mn,M} \Gamma^{mn}\delta_a^b -{1\over 8} e^m_M H_{mnp}\Gamma^{np} (\sigma^3)_{\; a}^{b}  \nonumber\\
& \quad + {1\over  8 \cdot 3!} e^{\phi} (F_{npq}- \chi H_{npq})\Gamma^{npq}  \Gamma_M (\sigma^1)_{\; a}^{b} \Big)
\end{align}

\subsection{Integrability of supersymmetry transformations}\label{apa1}

To examine the supersymmetries of of the background (\ref{planewavemetric}), (\ref{Tnfields}), we will first need to work out the spin connection. \begin{align}
X^\pm= {1\over 2} (T\pm X), \quad g_{+-}=g_{-+}=-2, \quad g^{+-}=g^{-+}= -{1\over 2}  
\end{align}
The frame forms fields    are given by
\begin{align}
e^i_\mu dx^\mu= dx^i, \quad e^+_\mu dx^\mu= dx^+, \quad e^-_\mu dx^\mu= dx^-+{1\over 4} \sum_k m_k x_k^2 dx^+
\end{align}
where $m_k=1$ for all $x^k$ except for $k=6$ for which  $m_6=3$. 
We can calculate the spin connection using the Cartan structure equations
\begin{align}
de^a+ \omega^a_{\; \;b} \wedge e^b=0
\end{align}
The the frame forms given above the only nontrivial one is
\begin{align}
de^-= {1\over 2} \sum_k m_k x_k dx^k\wedge dx^+
\end{align}
which implies that the connection 1-form is given by
\begin{align}
\omega^-_{\;\; k }={1\over 2}  m_k x^k dx^+
\end{align}
 and we find  that the only non vanishing components of the spin connection are given by
\begin{align}\label{spincon}
\omega_{+ k, +}= -m_k x^k
\end{align}
Note that because (\ref{spincon}) has two legs along the $x^+$ directions, the spin connection  will not contribute to the fermion action (\ref{Lrr}), since the  light-cone gauge $\Gamma^+\theta ^a = 0$ condition  will make it vanish.

In the following we will present the explicit form of the supersymmetry transformations in the string frame two component formalism following from (\ref{generalSUSYtransform}).
For the antisymmetric tensor fields of the $T_N$ example (\ref{Tnfields}) the dilatino variation takes the form\footnote{As discussed in section \ref{sec3} any Penrose limit can be brought into the $T_N$ form by a simple rotation in the 7-8 plane.}.
\begin{align} \label{dilavar}
\delta \lambda^a &= \Big(\Gamma^{67}(\sigma^3)^a_{\;b} +\Gamma^{68}(\sigma^1)^a_{\;b} \Big)\Gamma^+ \epsilon^b
\end{align}
and the gravitino variations are  given by
\begin{align}\label{gravivar}
\delta\psi_+^a &= \partial_+\epsilon^a+{1\over 2} \sum_k m_k x^k \Gamma^{k+}\epsilon^a+ \Gamma^{67}(\sigma^3)^a_{\;b} \epsilon^b-{1\over 2} \Gamma^{68}\Gamma^+\Gamma_+ (\sigma^1)^a_{\;b} \epsilon^b\nonumber \\
\delta\psi_i^a &=\partial_i \epsilon^a -{1\over 2}\Gamma^{68+}\Gamma_i  (\sigma^1)^a_{\;b} \epsilon^b, \quad i=1,2,3,4,5 \nonumber \\
\delta \psi_6^a &= \partial_6 \epsilon^a + \Gamma^{7+}(\sigma^3)^a_{\;b}  \epsilon^b -{1\over 2} \Gamma^{8+}(\sigma^1)^a_{\;b} \epsilon^b \nonumber \\
\delta \psi_7^a &= \partial_7 \epsilon^a -\Gamma^{6+} (\sigma^3)^a_{\;b} \epsilon^b   -{1\over 2} \Gamma^{678+}(\sigma^1)^a_{\;b} \epsilon^b  \nonumber \\
\delta \psi_8^a &= \partial_8 \epsilon^a +{1\over 2} \Gamma^{6+} (\sigma^1)^a_{\;b} \epsilon^b
 \end{align}
It was pointed out in \cite{Cvetic:2002nh} that pp waves have $16+N_{sup}$ Killing spinors; 16 of which must occur in any background while the remaining $0\leq N_{sup} \leq 16$ so-called "supernumerary" Killing spinors occur only in special backgrounds. After light-cone gauge fixing, only these extra spinors give rise to linearly-realized  supersymmetries. Such linearly realized supersymmetries also act as two dimensional world sheet supersymmetries implying a degeneracy of the world sheet energies of bosonic and fermionic excitations.   In our case, the sixteen $\epsilon$ which satisfy $\Gamma^+ \epsilon=0$ are the "automatic" supersymmetries. It's easy to check that for those the conditions (\ref{gravivar}) are integrable as only $\delta \psi_+$ is not automatically vanishing and can easily be integrated.

For the "supernumerary" Killing spinors the vanishing of the dilatino variation (\ref{dilavar}) imposes a projection condition on the supersymmetry transformation parameters $\epsilon$.  The integrability condition in the $i,+$ directions of $\delta\psi_M=0$ becomes
\begin{align}
(\partial_+ \partial_i - \partial_i \partial_+)\epsilon^a &= {1\over 2} \Gamma^{i +} \big(1-{1\over 2} \Gamma_+ \Gamma^+ \big) \epsilon^a\nonumber\\
&=  {1\over 4} \Gamma^{i +} \Gamma^+\Gamma_+\epsilon^a\nonumber \\
& =0
\end{align}
Where we used the identities
\begin{align}
\Gamma^+\Gamma_+= 1+\Gamma^{09}, \quad  \Gamma_+\Gamma^+= 1-\Gamma^{09}
\end{align}
Using the dilatino projection condition the $+, 7$ and $+,8$ integrability conditions are also satisfied. However the $+,6$ condition
takes the following form
\begin{align}
(\partial_+ \partial_6- \partial_6 \partial_+)\epsilon^a &= {1\over 2} \Gamma^{6+}  \big(\big(3 +{1\over 2} \Gamma_+ \Gamma^+\big)\delta^a_{\;b} - 2i\Gamma^{78}\big( 1+{1\over 2} \Gamma_+\Gamma^+\big) (\sigma^2)^a_{\;b} \big) \epsilon^b
\end{align}
Where the factor in parenthesis is not a projector and consequently there are no  "supernumerary"  Killing spinors in our background. This is to be contrasted with the Penrose limit of $AdS_5\times S^5$ where there are 16 additional supersymmetries \cite{Blau:2002dy}.

\section{Null geodesics}\label{appnull}
In this appendix we consider null geodesics in the warped $AdS_6\times S^2$ metric
\begin{align}\label{eqn:ansatzmetap}
	ds^2 &= f_6^2 (-\cosh \rho^2 dt + d\rho^2 +  \sinh\rho^2 ds_{S^4}^2)  + f_2^2 ( d\theta^2+\sin^2 \theta d\phi^2) 
	+ 4\hat \rho^2\, |dw|^2~,
	\end{align}

We consider geodesics which are located at the center of $AdS_6$ at  $\rho=0$ and  at the equator of  the two sphere  at $\theta={\pi \over 2}$. These choices preserve the  $U(1)$ of the two-sphere which we identify with a $U(1)$ inside the $SU(2)_R$ symmetry of the dual SCFT as well as the symmetries of the four-sphere of $AdS_6$, which means that the dual state has no angular momentum.

We consider however the possibility that the geodesic moves along the Riemann surface $\Sigma$, i.e. the coordinates $t, \phi, w$ and $\bar w$ depend on an affine parameter  $\lambda$ and the geodesic equation and null condition. With these choices  the geodesic equation

\begin{align}
{d^2 x^\mu \over  d\lambda^2}+ \Gamma^\mu_{\; \nu \rho} {d x^\nu\over d\lambda} {d x^\rho\over d\lambda} =0 
\end{align}
takes the following form
\begin{align}
{d\over d\lambda} \Big( {d t  \over d\lambda }    f^2_6(w , \bar w ) \Big)&=0   \label{geoa}\\
{d\over d\lambda} \Big( {d \phi \over d\lambda }   f^2_2(w, \bar w ) \Big)&=0 \label{geob} \\
{d^2 w  \over  d\lambda^2}- \left({d t  \over d\lambda }\right)^2 { \partial_{\bar w} f^2_6 \over 4 \hat \rho^2 }+\left({d \phi  \over d\lambda } \right)^2  {\partial_{\bar w} f^2_2 \over 4 \hat \rho^2 }+  \left( { d w  \over d\lambda }\right)^2{ \partial_w \hat \rho^2 \over \hat \rho^2}&=0  \\
{d^2 \bar w  \over  d\lambda^2}- \left({d t  \over d\lambda } \right)^2{ \partial_{ w} f^2_6 \over 4 \hat \rho^2 }+\left({d \phi  \over d\lambda }  \right)^2 {\partial_{w} f^2_2 \over 4 \hat \rho^2 }+  \left( { d \bar w  \over d\lambda }\right)^2{ \partial_{\bar w}\hat  \rho^2 \over  \hat  \rho^2}&=0\
\end{align}
Equations (\ref{geoa}) and (\ref{geob}) can be integrated  once
\begin{align}\label{tphigeo}
{d t  \over d\lambda } &=  {c_6 \over f_6^2} , \quad
{d \phi   \over d\lambda } =  {c_2 \over f_2^2} \
\end{align}
The condition that the geodesic is null $g_{\mu\nu}  {d x^\mu\over d\lambda} {d x^\nu\over d\lambda} =0$, becomes
\begin{align}
4 \hat \rho^2  {d w  \over d\lambda}{d \bar w  \over d\lambda} + {c_2^2 \over f_2^2} -{c_6^2\over f_6^2}=0
\end{align}

As discussed in section \ref{penroselim} our goal is to describe dual operators close to the BPS bound $\Delta=3J$ which leads to a condition on the coordinates $t$ and $\phi$ 
\begin{align}\label{bpscon}
{\partial \phi\over \partial t}=3
\end{align}
by identifying the generators of $t$ and $\phi$ translations with the scaling and R symmetry generators. Using (\ref{tphigeo}) this condition  implies
\begin{align}\label{geocona}
3&=  {c_2\over c_6} {f_6^2\over f_2^2}= 9 {c_2\over c_6} T^2   
\end{align}
Where (\ref{eq:metric-functions}) was used. It is easily confirmed that the choice used in the body of the paper, where $w$ is independent of $\lambda$ and fixed at the critical point  $w=w_c$, is indeed a null geodesic satisfying (\ref{bpscon})  for $c_6=3c_2$, due to the fact  that $T|_{w=w_c}=1$ and $\partial_w f_2|_{w=w_c}=\partial_w f_6|_{w=w_c}= 0$.

We briefly discuss the possibility of more general geodesics  where $w,\bar w$ depend on the affine parameter $\lambda$. The condition (\ref{geocona}) implies that  $T$ should be constant along such a geodesic. We have numerically searched for such geodesics for some examples and found that $T$ varies along the null geodesics. This provides evidence that one has to drop either the condition (\ref{bpscon}), which may make the field theory identification of the Penrose limit challenging or consider more general trajectories on $S^2$.  It would be very interesting to investigate the geodesics further and determine whether they are integrable. This  is  a quite challenging problem due to the complicated dependence of all the metric factors on the coordinates of $\Sigma$

We leave the investigation of these interesting   questions for future work.

\section{Plane wave limit for  $AdS_6$ solution of massive type IIA}
In \cite{Brandhuber:1999np} a solution of massive type IIA was found which is a warped product of $AdS_6$ over  part of a four sphere  $S^4$. This background was   constructed by taking a near horizon limit of a D4-D8 semi-localized brane solution. In this appendix we briefly compare the Penrose limit for this supergravity background \cite{Chong:2004kf} to the one for the type IIB $AdS_6$ solutions described in the body of this paper.
The metric and four form antisymmetric tensor field strength of the solution are given by
\begin{align}
ds^2&= {9\over 2} W(\xi)\Big( - \cosh^2 \rho dt^2+d\rho^2+\sinh^2\rho ds_{S^4}^2 +{4\over 9}\big( d\xi^2+ \sin^2\xi ds_{S_3}^2\big)\Big) \nonumber \\
F_4&= {20 \sqrt{2}\over 3} ( \cos \xi)^{1\over 3} \sin \xi^3 d\xi \wedge \omega_{S_3}, \quad \quad  e^{\phi} =(\cos \xi)^{-{5\over 6}}
\end{align}
Here the warp factor is given by $W(\xi)=\big(\cos \xi \big)^{1\over 6}$ in the string frame. The range of the warping coordinate $\xi\in [0, \pi/2]$ means that the three sphere warped over $\xi$ produces only  half of a four-sphere. The critical point of the warp factor of the three sphere is at $\xi=\pi/2$ and this corresponds to a Penrose limit considered in  \cite{Chong:2004kf},  which is analogous to the one discussed in our paper. There is however one important difference since $\xi=\pi/2$ corresponds to the $S^3$ along the equator of the $S^4$ and therefore to the strong coupling region where the supergravity approximation breaks down. In contrast, the critical point in the type IIB $AdS_6$ is a regular point where the supergravity approximation is valid for large $N$.  The same conclusion can be drawn from considering the T-dual of the  Brandhuber-Oz solution \cite{Lozano:2013oma} for which the holomorphic functions $\cA_{\pm}$ and $\cG$  are  \cite{DHoker:2016ujz}
\begin{align}
\cA_{\pm}= {a\over 2} w^2 \mp b w, \quad  \quad \cG= {ab\over 3} \big( 1+ (w+\bar w)^3\big)
\end{align}
Here the condition $\partial_w \cG=0$ is solved by $Im(w)=0$ which under T-duality is mapped to $\xi=\pi/2$ on the type IIA side.
\newpage
\providecommand{\href}[2]{#2}\begingroup\raggedright\endgroup

\end{document}